\documentclass[twocolumn,epjc3]{svjour3}

\usepackage{amsmath}
\usepackage{euscript,amssymb,epsfig}
\usepackage{graphicx}
\usepackage{amsfonts,latexsym}

\RequirePackage{graphicx}

\newcommand{\be}[1]{\begin{equation}\label{#1}}
\newcommand{\ee}{\end{equation}}
\newcommand{\ba}[1]{\begin{eqnarray}\label{#1}}
\newcommand{\ea}{\end{eqnarray}}
\newcommand{\rf}[1]{(\ref{#1})}
\newcommand{\nn}{\nonumber}

\journalname{Eur. Phys. J. C}

\begin{document}

\title{Rigorous theoretical constraint on constant negative EoS parameter $\omega$ and its effect for the late Universe}

\author{Alvina Burgazli\thanksref{e1,addr1} \and Maxim Eingorn\thanksref{e2,addr2,addr3,addr4} \and Alexander Zhuk\thanksref{e3,addr2}}

\thankstext{e1}{e-mail: aburgazli@gmail.com}

\thankstext{e2}{e-mail: maxim.eingorn@gmail.com}

\thankstext{e3}{e-mail: ai.zhuk2@gmail.com}

\institute{Department of Theoretical Physics, Odessa National University, Dvoryanskaya st. 2, Odessa 65082, Ukraine\\ \label{addr1} \and Astronomical
Observatory, Odessa National University, Dvoryanskaya st. 2, Odessa 65082, Ukraine\\ \label{addr2} \and Department of Theoretical and Experimental Nuclear
Physics, Odessa National Polytechnic University, Shevchenko av. 1, Odessa 65044, Ukraine\\ \label{addr3} \and CREST and NASA Research Centers, North Carolina
Central University, Fayetteville st. 1801, Durham, North Carolina 27707, U.S.A. \label{addr4}}

\date{Received: date / Accepted: date}

\maketitle

\begin{abstract}
In this paper, we consider the Universe at the late stage of its evolution and deep inside the cell of uniformity. At these scales, the Universe is filled with
inhomogeneously distributed discrete structures (galaxies, groups and clusters of galaxies). Supposing that the Universe contains also the cosmological
constant and a perfect fluid with a negative constant equation of state (EoS) parameter $\omega$ (e.g., quintessence, phantom or frustrated network of
topological defects), we investigate scalar perturbations of the FRW metrics due to inhomogeneities. Our analysis shows that, to be compatible with the theory
of scalar perturbations, this perfect fluid, first, should be clustered and, second, should have the equation of state parameter $\omega=-1/3$. In particular,
this value corresponds to the frustrated network of cosmic strings. Therefore, the frustrated network of domain walls with $\omega =-2/3$ is ruled out. A
perfect fluid with $\omega =-1/3$  neither accelerates nor decelerates the Universe. We also obtain the equation for the nonrelativistic gravitational
potential created by a system of inhomogeneities. Due to the perfect fluid with $\omega = -1/3$, the physically reasonable solutions take place for flat, open
and closed Universes. This perfect fluid is concentrated around the inhomogeneities  and results in screening of the gravitational potential. \end{abstract}


\section{\label{sec:1}Introduction}

\setcounter{equation}{0}

The accelerated expansion of the Universe at late stages of its evolution, found little more than a decade ago \cite{SN1,SN2}, is one of the most intriguing
puzzles of modern physics and cosmology. Recognition of this fact was the awarding of the Nobel Prize in 2011 to  Saul Perlmutter, Adam Riess and Brian
Schmidt. After their discovery, there were numerous attempts to explain the nature of such acceleration.
Unfortunately, there is no satisfactory explanation so far (see, e.g., the state of art in \cite{stateofart}). According to the recent observations
\cite{7WMAP,9WMAP,Planck2013}, the $\Lambda$CDM model is the preferable one. Here, the accelerated expansion is due to the cosmological constant. However, there is a
number of problems associated with the cosmological constant. Maybe, one of the main of them consists in the adjustment mechanism which could compensate
originally huge vacuum energy down to the cosmologically acceptable value and to solve the coincidence problem of close magnitudes of the non-compensated
remnants of vacuum energy and the energy density of the Universe at the present time \cite{Dolgov}. To resolve this problem, it was proposed to introduce
scalar fields as a matter source. Such scalar fields can be equivalently considered in the form of perfect fluids. Among these perfect fluids, a barotropic
fluid is one of the most popular objects of study. This fluid is characterized by the pressure which is the function of the energy density only: $p=p(\rho)$,
and the linear equation of state $p=\omega \rho$ is the most popular. These barotropic perfect fluids
with the equation of state parameters $\omega<-1/3$ can cause the accelerated expansion of the Universe. Such fluids are called quintessence
\cite{quintess1,quintess2,quintess3} and phantom \cite{phantom1,phantom2} for $-1<\omega<0$ and $\omega<-1$, respectively. Usually, they
have a time varying parameter $\omega$ of the equation of state. However, there is also a possibility to construct models with constant $\omega$ (for the
corresponding experimental restrictions see, particularly, Planck 2013 results \cite{Planck2013}). This imposes severe restrictions on the form of the scalar
field potential \cite{zhuk1996,ZBG}. In this case, a scalar field is equivalent to a perfect fluid with $\omega=\mbox{const}$. A large class of models is
expected to be well described (at least as far as the CMB anisotropy is concerned) by an effective constant equation of state parameter \cite{WL}. For example,
it is also well-known that frustrated networks of topological defects (cosmic strings and domain walls) have the form of perfect fluids with  the constant
parameter $\omega$ \cite{ZBG,ShellVil,Avelino,Kumar}. For example, $\omega=-1/3$ and $\omega=-2/3$ for cosmic strings and domain walls, respectively.
It is of interest to investigate the viability of the models with constant $\omega$ and to answer the question whether these models are an alternative to the
cosmological constant.

In our paper, we consider the compatibility of these models with the scalar perturbations of the Friedmann-Robertson-Walker (FRW) metrics. In the
hydrodynamical approach, such investigation was performed in a number of papers (see, e.g., \cite{Nov1} for $\omega =$ const and \cite{KS,Nov2} for
$\omega\neq$ const). We consider the Universe at late stages of its evolution
when galaxies and clusters of galaxies have already formed. At scales much larger than the characteristic distance between these inhomogeneities, the Universe
is well described by the homogeneous and isotropic FRW metrics. This is approximately 190 Mpc and larger \cite{EZcosm2}. At these scales, the matter fields
(e.g., cold dark matter) are well described by the hydrodynamical approach.  However, at smaller scales the Universe is highly inhomogeneous. Here, the
mechanical approach looks more adequate \cite{EZcosm2,EZcosm1}. It is worth noting that similar ideas concerning the discrete cosmology have been discussed in
the recent papers \cite{Ruth,EllGibb}. Obviously, at early stages of the Universe evolution (i.e. before the inhomogeneities formation when the density
contrast is much less than unity), the hydrodynamical approach works very well at small scales. It is clear that cosmological models should be tested at all
stages of the Universe evolution. It is not sufficient to show their compatibility with observations only at early stages, i.e. in the hydrodynamical approach,
as it was done in the previous papers. These models should also be in agreement with the mechanical approach. This is the aim and the novelty of our study. To
start with, in the present paper we consider the simplest model where a perfect fluid has a constant parameter of the equation of state. This article belongs
to a series of studies where we intend to test different cosmological models for their compatibility with the mechanical approach. Recently, such investigation
was performed for nonlinear $f(R)$ models \cite{ENZ1} as well as models with quark-gluon nuggets \cite{Laslo2}. In the following paper we will consider the
case of time-dependent parameters of the equation of state.

In mechanical approach, galaxies, dwarf galaxies and clusters of galaxies (all of them mostly composed of dark matter) can be considered as separate compact
objects. Moreover, at distances much greater than their characteristic sizes they can be well described as point-like matter sources. This is generalization of
the well-known astrophysical approach \cite{Landau} (see \S 106) to the case of dynamical cosmological background. Usually, gravitational fields of these
inhomogeneities are weak and their peculiar velocities are much less than the speed of light. Therefore, we can construct a theory of perturbations where the
considered point-like inhomogeneities disturb the FRW metrics. Such theory was proposed in the paper \cite{EZcosm1}. Then, we applied this mechanical approach
in \cite{EKZ2} to describe the mutual motion of galaxies, in particular, the Milky Way and Andromeda galaxies. For such investigations, the form of the
gravitational potential plays an important role. Hence, one of the main tasks of the present paper is to study a possibility to get a reasonable form of
gravitational potentials in the models with an additional perfect fluid with constant negative $\omega$. Then, if such potentials exist, we can study the
relative motion of galaxies in the field of these potentials and compare it with the corresponding motion in the $\Lambda$CDM model \cite{EKZ2}.

Because perfect fluids have $\omega=\mbox{const}$, their perturbations are purely adiabatic (see, e.g., \cite{BeanDore}), i.e. dissipative processes are
absent. Then, we demonstrate that, first, these fluids must be clustered (i.e. inhomogeneous) and, second, $\omega=-1/3$ is the only parameter which is
compatible with the theory of scalar perturbations. It is well known that such perfect fluid neither accelerates nor decelerates the Universe. Frustrated
network of cosmic strings is a possible candidate for such perfect fluid. It is worth noting that this conclusion is valid for perfect fluids with the constant
equation of state parameter. The conclusion for imperfect fluids (e.g., for scalar fields with arbitrary potentials) can be quite different.  We also obtain
formulas for the nonrelativistic gravitational potential created by a system of inhomogeneities (galaxies, groups and clusters of galaxies). We show that due
to the perfect fluid with $\omega=-1/3$, the physically reasonable expressions take place for flat, open and closed Universes.  If such perfect fluid is
absent, the hyperbolic space is preferred \cite{EZcosm1}. Hence, even if this perfect fluid does not accelerate the Universe, it can play an important role. It
is worth noting also that according to the paper \cite{Kumar}, a small contribution from the string network can explain the possible small departure from
$\Lambda$CDM evolution.

The paper is structured as follows. In Sec. 2, we consider scalar perturbations in the Friedmann Universe filled with the cosmological constant, pressureless
dustlike matter (baryon and dark matter) and perfect fluid with negative constant equation of state. Here, we get the equation for the nonrelativistic
gravitational potential. In Sec. 3, we find solutions of this equation for an arbitrary system of inhomogeneities for flat, open and closed Universes. These
solutions have the Newtonian limit in the vicinity of inhomogeneities and are finite at any point outside inhomogeneities. The main results are summarized in
concluding Sec. 4.


\section{\label{sec:2}Scalar perturbations of FRW Universe }

\setcounter{equation}{0}

{\it Homogeneous background.}

To start with, we consider a homogeneous isotropic Universe described by the FRW metrics
\ba{2.1} ds^2&=&a^2\left(d\eta^2-\gamma_{\alpha\beta}dx^{\alpha}dx^{\beta}\right)\nn\\
&=&a^2\left(d\eta^2-d\chi^2 -\Sigma^2(\chi ) d\Omega^2_2\right)\, , \ea
where
\be{2.1a}
\Sigma (\chi)=\left\{
\begin{array}{ccc}
\sin \chi\, ,  & \chi \in [0,\pi] & \mbox{for} \quad {\mathcal K}=+1 \\
\chi\, ,  & \chi \in [0,+\infty) & \mbox{for} \quad {\mathcal K}=0 \\
\sinh \chi\, ,  & \chi \in [0,+\infty) & \mbox{for} \quad {\mathcal K}=-1
\end{array} \right.
\ee
and $\mathcal K=-1,0,+1$ for open, flat and closed Universes, respectively. As matter sources, we consider the cosmological constant\footnote{Perfect fluids
(e.g., quintessence and phantom) with the negative parameter of the equation of state $\omega<-1/3$ were introduced to explain the late time acceleration of
the Universe. They are an alternative to the cosmological constant. However, in our model, we shall keep both perfect fluids and the cosmological constant
because we investigate the full range of negative parameters $\omega<0$. Moreover, we shall show that the only possible value of $\omega$ for the considered
perfect fluid is $-1/3$. Then, the inclusion of $\Lambda$ becomes justified. Additionally, a small contribution from these fluids (e.g., frustrated network of
cosmic strings with $\omega=-1/3$) can explain the possible small departure from $\Lambda$CDM evolution \cite{Kumar}.} $\Lambda$, pressureless dustlike matter
(in accordance with the current observations \cite{7WMAP,9WMAP}, we assume that dark matter (DM) gives the main contribution to this matter) and an additional
perfect fluid with the equation of state $\overline p=\omega\overline \varepsilon$ where $\omega<0$. In the present paper, $\omega =\mbox{const}$. As we
already wrote in the introduction, such perfect fluids can be modeled by scalar fields with the corresponding form of the potentials \cite{zhuk1996,ZBG} as
well as by the frustrated network of the topological defects \cite{ZBG,ShellVil,Avelino,Kumar}.  We exclude the values $\omega =0,-1$ because they are
equivalent to DM and the cosmological constant, respectively. Scalar fields with $-1<\omega<0$ and $\omega<-1$ are usually called quintessence and phantom,
respectively. Below, the overline denotes homogeneous perfect fluids. It can be easily seen from the conservation equation that in the case of the homogeneous
perfect fluid
\be{2.2}
\overline \varepsilon=\varepsilon_0\frac{a_0^{3(1+\omega)}}{a^{3(1+\omega)}}\, ,
\ee
where $a_0$ is the scale factor at the present time and $\varepsilon_0$ is the current value of the energy density $\overline{\varepsilon}$.

Because we consider the late stages of the Universe evolution, we neglect the contribution of radiation. It is worth noting that radiation can be also included
into consideration \cite{EZcosm2}, and the simple analysis demonstrates that this does not affect the results of the paper. Therefore, the Friedmann equations
read
\be{2.3} \frac{3\left(\mathcal{H}^2+\mathcal{K}\right)}{a^2}=\kappa\overline{T}^0_0+\Lambda+\kappa\overline \varepsilon \ee
and
\be{2.4}
\frac{2\mathcal{H}'+\mathcal{H}^2+\mathcal{K}}{a^2}=\Lambda-\kappa\omega\overline \varepsilon\, ,
\ee
where ${\mathcal H}\equiv a'/a\equiv (da/d\eta)/a$ and $\kappa\equiv 8\pi G_N/c^4$ ($c$ is the speed of light and $G_N$ is the Newton's gravitational
constant). Here, $\overline T^{ik}$ is the energy-momentum tensor of the average pressureless dustlike matter. For such matter, the energy density $\overline
T^{0}_{0} =\overline \rho c^2/a^3$ is the only nonzero component. $\overline \rho=\mbox{const}$ is the comoving average rest mass density \cite{EZcosm1}. It is
worth noting that in the case $\mathcal K =0$ the comoving coordinates $x^{\alpha}$ may have a dimension of length, but then the conformal factor $a$ is
dimensionless, and vice versa. However, in the cases $\mathcal K=\pm 1$ the dimension of $a$ is fixed. Here, $a$ has a dimension of length and $x^{\alpha}$ are
dimensionless. For consistency, we shall follow this definition for $\mathcal K=0$ as well. For such choice of the dimension of $a$, the rest mass density has
a dimension of mass.


Conformal time $\eta$ and synchronous time $t$ are connected as $cdt=a d\eta$. Therefore, eqs. \rf{2.3} and \rf{2.4}, respectively, take the form
\be{2.4a} H^2=H_0^2 \left(\Omega_{M}\frac{a_0^3}{a^3}+\Omega_{\Lambda}+\Omega_{\mathcal K}\frac{a_0^2}{a^2}
+\Omega_{\mathrm{pf}}\frac{a_0^{3(1+\omega)}}{a^{3(1+\omega)}}\right)\ee
and
\be{2.5}
\frac{\ddot a}{a}=H_0^2\left(-\frac{1}{2}\Omega_{M}\frac{a_0^3}{a^3}+\Omega_{\Lambda} -\frac12 (1+3\omega)
\Omega_{\mathrm{pf}}\frac{a_0^{3(1+\omega)}}{a^{3(1+\omega)}}\right)\, ,
\ee
where $a_0$ and $H_0$ are the values of the conformal factor $a$ and the Hubble "constant" $H\equiv \dot a/a\equiv (da/dt)/a$ at the present time $t=t_0$, and
we introduced the density parameters:
\ba{2.5a} &{}&\Omega_M=\frac{\kappa\overline\rho c^4}{3H_0^2a_0^3},\quad \Omega_{\Lambda}=\frac{\Lambda c^2}{3H_0^2}\, ,\quad \nn\\
&{}&\Omega_{\mathcal K}= -\frac{\mathcal Kc^2}{a_0^2H_0^2},\quad \Omega_{\mathrm{pf}}=\frac{\kappa c^2 \varepsilon_0}{3H_0^2}\, ,\ea
therefore
\be{2.5b}
\Omega_M+\Omega_{\Lambda}+\Omega_{\mathcal K}+ \Omega_{\mathrm{pf}}=1\, .
\ee
It is of interest to get the experimental restriction on $\Omega_{\mathrm{pf}}$. This requires a separate investigation which is out of the scope of our paper.
We can easily see from Eq. \rf{2.5} that perfect fluids with $\omega <-1/3$ can provide the accelerated expansion of the Universe.

\

{\it Scalar perturbations.}

As we have written in the Introduction, the inhomogeneities in the Universe result in scalar perturbations of the metrics \rf{2.1}. In the conformal Newtonian
gauge, such perturbed metrics is \cite{Mukhanov-book,Rubakov-book}
\be{2.6}
ds^2\approx a^2\left[(1+2\Phi)d\eta^2-(1-2\Psi)\gamma_{\alpha\beta}dx^{\alpha}dx^{\beta}\right]\, ,
\ee
where scalar perturbations $\Phi,\Psi \ll 1$. Following the standard argumentation, we can put $\Phi=\Psi$. We consider the Universe at the late stage of its
evolution when the peculiar velocities of inhomogeneities (both for dustlike matter and the considered perfect fluid) are much less than the speed of light:
\be{2.7}
\frac{dx^{\alpha}}{d\eta}
=a\frac{dx^{\alpha}}{dt} \frac{1}{c}\equiv \frac{v^{\alpha}}{c}\ll 1\, .
\ee
We should stress that smallness of the nonrelativistic gravitational potential $\Phi$ and peculiar velocities $v^{\alpha}$ are two independent conditions
(e.g., for very light relativistic masses the gravitational potential can still remain small). Under these conditions, the gravitational potential $\Phi$
satisfies the following system of equations (see \cite{EZcosm1} for details\footnote{\label{linear}It is well known that in the hydrodynamic approach, the
linear formalism is not applicable to study the formation of galaxies and clusters of galaxies. However, first, we consider the late stage of the Universe
evolution when these inhomogeneities were mainly formed. Second, in our mechanical approach, we can use the linear approximation due to the smallness of the
gravitational fields and peculiar velocities. Here, the structure of the galaxies can evolve on account of mechanical merger of inhomogeneities.}):
\ba{2.8} &{}&\Delta\Phi-3\mathcal{H}(\Phi'+\mathcal{H}\Phi)+3\mathcal{K}\Phi=\frac{1}{2}\kappa a^2\delta T_0^0+\frac{1}{2}\kappa a^2\delta\varepsilon\, ,\nn\\
&{}&\\
\label{2.9}
&{}&\frac{\partial}{\partial x^\beta}(\Phi'+\mathcal{H}\Phi)=0\, ,\\
\label{2.10} &{}&\Phi''+3\mathcal{H}\Phi'+(2\mathcal{H}'+\mathcal{H}^2)\Phi-\mathcal{K}\Phi=\frac{1}{2}\kappa a^2\delta p\, , \ea
where the Laplace operator
\be{2.8a}
\triangle=\frac{1}{\sqrt{\gamma}}\frac{\partial}{\partial x^{\alpha}}\left(\sqrt{\gamma}\gamma^{\alpha\beta}\frac{\partial}{\partial x^{\beta}}\right)
\ee
and $\gamma$ is the determinant of $\gamma_{\alpha\beta}$. Following the reasoning of \cite{EZcosm1}, we took into account that peculiar velocities of
inhomogeneities are nonrelativistic, and under the corresponding condition \rf{2.7} the contribution of $\delta T_{\beta}^0$ is negligible compared to that of
$\delta T_{0}^0$ both for dustlike matter and the considered perfect fluid. Really, according to \cite{EZcosm1}, the true rest mass density $\rho$ of usual
matter, presented by a sum of delta-functions (see Eq. \rf{3.4} below), is comparable with itself after subtracting the average value $\overline\rho$.
Consequently, $\delta T_{\beta}^0/\delta T_0^0\sim v^{\beta}/c\ll1$. Exactly the same strong inequality holds true also for the additional perfect fluid under
the quite natural assumption that only its fraction of the order $\delta\varepsilon/\varepsilon$ takes part in considerable motion due to interaction between
inhomogeneities.
In other words, account of $\delta T_{\beta}^0$ is beyond the accuracy of the model. This approach is completely consistent with \cite{Landau} where it is
shown that the nonrelativistic gravitational potential is defined by the positions of the inhomogeneities but not by their velocities (see Eq. (106.11) in this
book). In the case of an arbitrary number of dimensions, a similar result was obtained in \cite{EZ3}. On the other hand, the motion of nonrelativistic
inhomogeneities is defined by the gravitational potential (see, e.g., \cite{EKZ2}). The perturbed matter remains nonrelativistic (pressureless) that results in
the condition $\delta T_{\beta}^{\alpha}=0$. For the considered perfect fluid we have $\delta T_{\beta}^{\alpha}=-\delta p\delta_{\beta}^{\alpha}$, and
$\delta\varepsilon$ is a fluctuation of the energy density for this perfect fluid. In \rf{2.8} $\delta T^0_0$ is related to the fluctuation of the energy
density of dustlike matter and has the form \cite{EZcosm1}:
\be{2.11}
\delta T_{0}^0=\frac{\delta\rho c^2}{a^3}+\frac{3\overline{\rho}c^2\Phi}{a^3}\, ,
\ee
where $\delta\rho$ is the difference between real and average rest mass densities:
\be{2.12}
\delta\rho = \rho-\overline\rho\, .
\ee
From Eq. \rf{2.9} we get
\be{2.13}
\Phi(\eta,{\bf r})=\frac{\varphi({\bf r})}{c^2a(\eta)}\, ,
\ee
where $\varphi({\bf r})$ is a function of all spatial coordinates and we have introduced $c^2$ in the denominator for convenience. Below, we shall see that
$\varphi({\bf r})\sim 1/r$ in the vicinity of an inhomogeneity, and the nonrelativistic gravitational potential $\Phi(\eta,{\bf r})\sim 1/(a r)=1/R$, where
$R=ar$ is the physical distance. Hence, $\Phi$ has the correct Newtonian limit near the inhomogeneities. Substituting the expression \rf{2.13} into Eqs.
\rf{2.8} and \rf{2.10}, we get the following system of equations:
\ba{2.14}
&{}&\frac{1}{a^3}\left(\Delta\varphi+3\mathcal{K}\varphi\right)=\frac{1}{2}\kappa c^2\delta T_0^0+\frac{1}{2}\kappa c^2\delta\varepsilon\, ,\\
&{}&\label{2.15}
\frac{1}{a^3}(\mathcal{H}'-\mathcal{H}^2-\mathcal{K})\varphi=\frac{1}{2}\kappa c^2\delta p\, .
\ea
From the Friedmann equations \rf{2.3} and \rf{2.4} we obtain
\be{2.16}
\frac{1}{a^3}\left(\mathcal{H}'-\mathcal{H}^2-\mathcal{K}\right)
=\frac{1}{2a}\left(-\kappa\overline{T}_0^0-\kappa(1+\omega)\overline \varepsilon\right)\, .
\ee
Then, Eq. \rf{2.15} reads
\be{2.17} \left(-\kappa\frac{\overline{\rho}c^2}{a^4}-\kappa(1+\omega)\frac{\varepsilon_0}{a_0}\frac{a_0^{4+3\omega}}{a^{4+3\omega}}\right)\varphi=\kappa
c^2\omega\delta\varepsilon\, .\ee
It should be noted that we consider the perfect fluids without thermal coupling to any other type of matter. It means, in particular, that evolution of its
homogeneous background as well as scalar perturbations occurs adiabatically or, in other words, without change of entropy. Therefore, in the case of the
constant parameter of the equation of state we preserve the same linear equation of state $\delta p=\omega\delta\varepsilon$ with the same constant parameter
$\omega$ for the scalar perturbations $\delta p$ and $\delta\varepsilon$ of pressure and energy density respectively, as for their background values $\overline
p$ and $\overline\varepsilon$ (see, e.g., equations (1) and (2) in \cite{BeanDore}). Obviously, imperfect fluids such as scalar fields with arbitrary
potentials (which results in time-dependent parameter $\omega$) require a different approach \cite{imperfect1,imperfect2,imperfect3,imperfect4}.

Taking into account the expression \rf{2.13}, we get that in the right hand side of Eq. \rf{2.11} the second term is proportional to $1/a^4$ and should be
dropped because we consider the nonrelativistic matter{\footnote{Radiation can be easily included in our scheme \cite{EZcosm2}. The simple analysis shows that
this does not change all of the following results.}. This is the accuracy of our approach, i.e. for the terms of the form of $1/a^n$, we drop ones with $n\ge
4$ and leave terms with $n<4$. Obviously, $4+3\omega <4$ for $\omega <0$.
Hence, we can draw the important conclusion regarding the purely homogeneous non-clustered quintessence/phantom fluids with $\delta p,\delta \varepsilon =0$.
For these fluids, we arrive at a contradiction because in Eq. \rf{2.17} the right hand side is equal to zero while the left hand side is nonzero. Therefore,
such fluids are forbidden{\footnote{It can be easily realized that the homogeneous solution $\delta\varepsilon =0$, $\varphi =0$ is forbidden because it
contradicts Eq. \rf{2.14}. The point is that the standard matter density perturbations $\delta T^0_0$ defined in Eq. \rf{2.11} are supposed to be nonzero. In
other words, we consider the Universe filled with inhomogeneously distributed galaxies, groups and clusters of galaxies. The presence of these inhomogeneities
results in nonzero perturbations of the 00 component of the corresponding energy-momentum tensor \cite{EZcosm1}.}}. The considered perfect fluid (quintessence,
phantom or frustrated network of topological defects) should be capable of clustering.
In the papers \cite{quintess3,SWZ}, it was also pointed out that the quintessence has to be inhomogeneous. For the inhomogeneous perfect fluid we get from Eq.
\rf{2.17} that
\be{2.18}
\delta\varepsilon=
-\frac{1+\omega}{c^2\omega}\varepsilon_0\frac{a_0^{3+3\omega}}{a^{4+3\omega}}\varphi\, .
\ee
Substituting \rf{2.18} into \rf{2.14}, we obtain within our accuracy
$$
\frac{1}{a^3}\left(\Delta\varphi+3\mathcal{K}\varphi\right)=\frac{1}{2}\kappa c^2\frac{\delta\rho c^2}{a^3}-\frac{1}{2}\kappa
c^2\frac{1+\omega}{c^2\omega}\varepsilon_0\frac{a_0^{3+3\omega}}{a^{4+3\omega}}\varphi
$$
\be{2.19}
\Rightarrow  \Delta\varphi+3\mathcal{K}\varphi=\frac{1}{2}\kappa c^4\delta\rho -
\frac{1+\omega}{2\omega}\kappa\varepsilon_0 a_0^2\frac{a_0^{1+3\omega}}{a^{1+3\omega}}\varphi\, .
\ee
In this equation, all terms except the last one do not depend on time{\footnote{We would like to remind that quantities $\varphi$ and $\delta\rho$ are the
comoving ones \cite{EZcosm1}. Therefore, within the adopted accuracy when both nonrelativistic and weak field limits are applied, they do not depend explicitly on time \cite{EZcosm2}.}}. Therefore, $\omega =-1/3$ is the only possibility to avoid this problem.
Hence, we arrive at the following important conclusion. At the late stage of the Universe evolution, the considered perfect fluids are compatible with the
scalar perturbations only if, first, they are inhomogeneous, and, second, they have the equation of state parameter $\omega=-1/3$. For example, the frustrated
network of cosmic strings can be a candidate for this fluid. On the other hand, frustrated domain walls are ruled out because they have $\omega=-2/3$. Eq.
\rf{2.5} clearly demonstrates that the perfect fluid with $\omega=-1/3$ neither accelerates nor decelerates the  Universe.

It is worth noting that in our model neither the nonrelativistic gravitational potential $\Phi \sim 1/a$ nor the perfect fluid density contrast
$\delta\varepsilon/\overline \varepsilon \sim 1/a$ diverge with time (with the scale factor $a$) in spite of the negative sign of the ratio $\delta p/\delta
\varepsilon$ which is often treated as the speed of sound squared. In the papers \cite{BS,BCCM} it was shown that such components could be stable if
sufficiently rigid. Really, as we shall show below, our perfect fluid is not purely fluid. Its fluctuations are concentrated around the matter/dark matter
inhomogeneities (see, e.g., Eq. \rf{3.8}). Obviously, the speed of sound in this case is close to zero. As noted in the paper \cite{Conversi}, for the "solid"
dark energy, the zero speed of sound is preferable. On the other hand, due to the concentration of fluctuations around the matter/dark matter inhomogeneities,
they have velocities of the order of the velocities of matter/dark matter. That is, the condition \rf{2.7} is valid for the perfect fluid in spite of the
averaged relativistic equation of state $\overline p =\omega\overline\rho$.

For $\omega=-1/3$, the equation for the gravitational potential and the fluctuation of the energy density of the perfect fluid read, respectively:
\be{2.20}
\Delta\varphi+\left(3\mathcal{K}-\frac{8\pi G_N}{c^4}\varepsilon_0a_0^2\right)\varphi=4\pi G_N
(\rho-\overline{\rho})\,
\ee
and
\be{2.21}
\delta\varepsilon=\frac{2\varepsilon_0a_0^2}{c^2a^3}\varphi\, .
\ee
Naturally, Eq. \rf{2.20} coincides with the Eq. (2.27) in \cite{EZcosm1} in the absence of the perfect fluid (i.e. for $\varepsilon_0=0$). Moreover, for
$\mathcal{K}=0$ and $\varepsilon_0=0$, this equation coincides (up to evident redefinition) with Eq. (7.14) in the well-known book \cite{Peebles} and Eq. (2)
for the GADGET-2 \cite{Gadget2}.

In the next section, we shall investigate Eq. \rf{2.20} depending on the curvature parameter $\mathcal{K}$. We shall show that reasonable expressions of the
conformal gravitational potential $\varphi$ exist for any sign of $\mathcal{K}$. This takes place due to the presence of the perfect fluid with $\omega =-1/3$.
If this fluid is absent, the hyperbolic model $\mathcal{K}=-1$ is preferred \cite{EZcosm1}. Therefore, the positive role of such perfect fluid is that its
presence gives a possibility to consider models for any $\mathcal{K}$.


\section{\label{sec:3}Gravitational potentials}

\setcounter{equation}{0}

It is convenient to rewrite Eq. \rf{2.20} as follows:
\be{3.1}
\Delta \phi -\lambda^2\phi= 4\pi G_N\rho \, ,
\ee
where the truncated gravitational potential is
\be{3.2} \phi = \varphi - \frac{4\pi G_N\overline\rho}{\lambda^2}\, ,\quad \lambda \neq 0\, , \ee
and
\be{3.3}
\lambda^2\equiv \frac{8\pi G_N}{c^4}\varepsilon_0a_0^2-3 \mathcal{K}=\frac{3a_0^2H_0^2}{c^2} \left(\Omega_{\mathcal K}+ \Omega_{\mathrm{pf}}\right)\, .
\ee
As we have already mentioned in the Introduction, on scales smaller than the cell of uniformity size and on late stages of evolution, the Universe is filled
with inhomogeneously distributed discrete structures (galaxies, groups and clusters of galaxies) with dark matter concentrated around these structures. Then,
the rest mass density $\rho$ reads \cite{EZcosm1}
\be{3.4}
\rho=\frac{1}{\sqrt{\gamma}}\sum_i m_i \delta({\bf r}-{\bf r}_i)\, ,
\ee
where $m_i$ is the mass of $i-$th inhomogeneity. Therefore, Eq. \rf{3.1} satisfies the very important principle of superposition. It is sufficient to solve
this equation for one gravitating mass $m_i$ and obtain its gravitational potential $\phi_i$. The gravitational potential for all system of inhomogeneities is
equal to a sum of potentials $\phi_i$. It is worth recalling that the operator $\Delta$ is defined by Eq. \rf{2.8a}. As boundary conditions, we demand that,
first, the gravitational potential of a gravitating mass should have the Newtonian limit near this inhomogeneity $\phi_i\sim 1/r$ and, second, this potential
should converge at any point of the Universe (except the gravitating mass position).

It seems reasonable to assume also that the total gravitational potential
averaged over the whole Universe is equal to zero (see, e.g., \cite{EZcosm1}):
\be{3.5}
\overline \varphi = \overline\phi +\frac{4\pi G_N\overline\rho}{\lambda^2}=0\, ,\quad \overline \phi =
\sum_i\frac{1}{V}\int_{V}\phi_i dV\, ,
\ee
where $V$ is the volume of the Universe. This demand results in another physically reasonable condition: $\overline {\delta\varepsilon} =0$ (see Eq. \rf{2.21}).

\vspace{0.5cm}

{\it Flat space: $\mathcal{K}=0$.}

\vspace{0.5cm}

In the case $\varepsilon_0>0 \to \lambda^2 = \frac{8\pi G_N}{c^4}\varepsilon_0a_0^2 >0$, the solution of \rf{3.1} for a separate mass $m_i$ satisfying the
mentioned above boundary conditions reads
\be{3.6}
\phi_i=-\frac{G_Nm_i}{r}\exp(-\lambda r)\, , \quad \lambda>0\, , \; 0<r<+\infty\, .
\ee
It can be easily seen that this truncated potential has the Newtonian limit for $r\to 0$. This expression shows that the perfect fluid results in the screening
of the Newtonian potential. A similar effect for the Coulomb potential takes place in plasma. In our case, the screening originates due to specific nature of
the perfect fluid. It is worth mentioning that the exponential screening of the gravitational potential was introduced "by hand" in a number of models to solve
the famous Seeliger paradox (see, e.g., the review \cite{Norton}). In our model, we resolve this paradox in a natural way due to the presence of the specific
perfect fluid.

For a many-particle system, the total gravitational potential takes the form
\be{3.7}
\varphi=-G_N\sum_i\frac{m_i}{|{\bf r}-{\bf r}_i|}\exp\left(-\lambda |{\bf r}-{\bf r}_i|\right)+\frac{4\pi G_N\overline\rho}{\lambda^2}\, .
\ee
Substituting \rf{3.7} into \rf{2.21}, we get for the fluctuations of the perfect fluid energy density the following expression:
\be{3.8} \delta\varepsilon=\frac{2\varepsilon_0a_0^2}{c^2a^{3}}\left(-G_N\sum_im_i\frac{\exp\left(-\lambda |{\bf r}-{\bf r}_i|\right)}{|{\bf r}-{\bf r}_i|}+
\frac{c^4\overline{\rho}}{2\varepsilon_0a_0^{2}}\right)\,  \ee
Therefore, we arrive at a physically reasonable conclusion that these fluctuations are concentrated around the matter/dark matter inhomogeneities and the
corresponding profile is given by Eq. \rf{3.8}.

The averaged value of the $i$-th component of the truncated potential over some finite volume $V_0$ is
\ba{3.9}
\overline \phi_i &=& \frac{4\pi}{V_0}\int_{0}^{r_0}\left[-G_N m_i\frac{\exp (-\lambda r)}{r}\right]r^2dr\nn\\
&=&-\frac{4\pi G_N m_i}{V_0}\left[-\frac{\exp(-\lambda r_0)}{\lambda}\left(r_0+\frac{1}{\lambda}\right)+\frac{1}{\lambda^2}\right]\, . \ea
Then, letting the volume go to infinity ($r_0\to +\infty \Rightarrow V_0\to +\infty$) and taking all gravitating masses, we obtain
\be{3.10}
\overline \phi =-G_N\overline{\rho}\frac{4\pi}{\lambda^2}\, ,
\ee
where $\overline\rho = \lim\limits_{V_0\to +\infty}\sum\limits_i m_{i}/V_0$.  Therefore, the averaged gravitational potential \rf{3.5} is equal to zero:
$\overline\varphi=0$.
Consequently, $\overline {\delta\varepsilon} =0$.

The case $\varepsilon_0<0 \Rightarrow \lambda^2 \equiv -\mu^2 <0$ is not of interest. Here, we get the expression $\phi_i = -(G_Nm_i/r)\cos(\mu r)$ which does
not have clear physical sense. Additionally, this expression does not allow the procedure of averaging.

\vspace{0.5cm}

{\it Spherical space: $\mathcal{K}=+1$.}

\vspace{0.5cm}

Let us consider, first, the case $\lambda^2 = \frac{8\pi G_N}{c^4}\varepsilon_0a_0^2-3\equiv -\mu^2 <0$. This case is of interest because it allows us to
perform the transition to small values of the energy density of the perfect fluid: $\varepsilon_0\to 0$. Here, the solution of \rf{3.1} for a separate mass
$m_i$ is
\be{3.11}
\phi_i=-G_Nm_i\frac{\sin\left[(\pi-\chi)\sqrt{\mu^2+1}\, \right]}{\sin\left(\pi\sqrt{\mu^2+1}\, \right)\sin{\chi}}\, , \quad 0<\chi \leq \pi\, .
\ee
For $\sqrt{\mu^2+1} \neq 2,3,\ldots$ (we would remind that $\mu^2\neq 0$), this formula is finite at any point $\chi \in (0,\pi]$ and has the Newtonian limit
for $\chi\to 0$. In the case of absence of the perfect fluid $\varepsilon_0=0 \to \sqrt{\mu^2+1} = 2$, this expression is divergent at $\chi=\pi$. We
demonstrated this fact in our paper \cite{EZcosm1}. Therefore, the considered perfect fluid gives a possibility to avoid this problem for the models with
$\mathcal{K}=+1$. It can be easily verified that for the total system of gravitating masses, the averaged value of the total truncated potential has the form
of \rf{3.10} that results in $\overline \varphi =0 \Rightarrow \overline{\delta \varepsilon}=0$.

In the case $\lambda^2>0$, the formulas can be easily found from \rf{3.11} with the help of analytical continuation $\mu \to i\mu$. In other words, it is
sufficient in Eq. \rf{3.11} to replace $\mu^2$ by $-\lambda^2$. The obtained expression is finite for all $\chi \in (0,\pi]$ and the averaged gravitational
potential is equal to zero: $\overline \varphi =0 \Rightarrow \overline{\delta \varepsilon}=0$.

\vspace{0.5cm}

{\it Hyperbolic space: $\mathcal{K}=-1$.}

\vspace{0.5cm}

Here, the most interesting case corresponds to  $\lambda^2 = \frac{8\pi G_N}{c^4}\varepsilon_0a_0^2+3 > 0$. This choice of sign gives a possibility to perform
the transition to small values of the energy density of the perfect fluid: $\varepsilon_0\to 0$. Then, the desired solution of Eq. \rf{3.1} for a mass $m_i$ is
\be{3.12}
\phi_i=-\frac{G_Nm_i}{\sinh{\chi}}\exp\left(-\chi\sqrt{\lambda^2+1}\right)\, , \quad 0<\chi <+\infty \, .
\ee
If the perfect fluid is absent ($\varepsilon_0=0$), then we reproduce the formula obtained in \cite{EZcosm1}. On the other hand, the expression \rf{3.12} shows
that for $\varepsilon_0 >0\to \lambda^2 +1 > 4$, the perfect fluid enhances the screening of the gravitating mass. For a many-particle system, the total
gravitational potential takes the form
\be{3.13}
\varphi=-G_N\sum_i m_i\frac{\exp(-l_i\sqrt{\lambda^2+1}\, )}{\sinh l_i}+\frac{4\pi G_N\overline\rho}{\lambda^2}\, ,
\ee
where $l_i$ denotes the geodesic distance between the $i$-th mass $m_i$ and the point of observation. Similarly, using Eq. \rf{3.11}, we can write the
expression for the total potential in the case of the spherical space.

Taking into account that the averaged total truncated potential has again the form \rf{3.10}, the procedure of averaging leads to the physically reasonable
result: $\overline \varphi =0 \Rightarrow \overline{\delta \varepsilon}=0$.

Concerning the case $\lambda^2<0$, the truncated gravitational potential is finite in the limit $\chi\to +\infty $. However, the procedure of averaging does
not exist here. Therefore, this case is not of interest for us.

To conclude this section, we discuss briefly the case $\lambda^2=0$.
For $\mathcal{K}=0,-1$, the principle of superposition is absent now. To make the gravitational potential finite at any point including the spatial infinity,
we need to cutoff it smoothly at some distances from each gravitating mass. If $\mathcal{K}=0$, then the perfect fluid is absent and this case was described in
detail in \cite{EZcosm1}. It was shown that the averaged gravitational potential is not equal to zero. This is a disadvantage of such models. In the case
$\mathcal{K}=+1$, the principle of superposition can be introduced due to the finiteness of the total volume of the Universe. Here, the comoving averaged rest
mass density can be split as follows: $\overline \rho = \sum_i m_i/(2\pi^2) \equiv \sum_i\overline \rho_i$. Then, Eq. \rf{2.20} can be solved separately for
each combination $(m_i,\overline \rho_i)$. As a result, the gravitational potential of the $i$-th mass is
\be{3.14}
\varphi_i=\frac{G_Nm_i}{2\pi}-G_Nm_i\frac{\cos{\chi}}{\sin{\chi}}\left(1-\frac{\chi}{\pi}\right)\, ,\quad 0<\chi \leq \pi\, .
\ee
This potential is convergent at any point $\chi\neq 0$, including $\chi=\pi$. It is not difficult to see that $\overline \varphi_i=0$. Therefore, the total
averaged gravitational potential is also equal to zero: $\overline \varphi=\sum_i\overline\varphi_i =0 \Rightarrow \overline {\delta \varepsilon}=0$.


\section{\label{sec:4}Conclusion}

In our paper, we have considered the perfect fluids with the constant negative parameter $\omega$ of the equation of state. We have investigated the role of
these fluids  for the Universe at late stages of its evolution. Such perfect fluids can be simulated by scalar fields with the corresponding form of the
potentials \cite{zhuk1996,ZBG} as well as by the frustrated network of the topological defects \cite{ZBG,ShellVil,Avelino,Kumar}. Scalar fields with
$-1<\omega<0$ and $\omega<-1$ are usually called quintessence and phantom, respectively, and they can be an alternative to the cosmological constant explaining
the late time acceleration of the Universe. It takes place if their parameter of the equation of state $\omega<-1/3$. On the other hand, a small contribution
from these fluids (e.g., the frustrated network of cosmic strings with $\omega=-1/3$) can explain the possible small departure from $\Lambda$CDM evolution
\cite{Kumar}.

To check the compatibility of these fluids with observations, we considered the present Universe at scales much less than the cell of homogeneity size which is
approximately $190$ Mpc \cite{EZcosm2}. At such distances, our Universe is highly inhomogeneous and the averaged Friedmann approach does not work here. We need
to take into account the inhomogeneities in the form of galaxies, groups and clusters of galaxies. It is natural to assume also that the perfect fluid
fluctuates around its average value. Therefore, these fluctuations as well as inhomogeneities perturb the FRW metrics. To consider these perturbations inside
the cell of uniformity, we need to use the mechanical approach. This approach was established in our papers \cite{EZcosm2,EZcosm1}. This is the novelty of our
present work because in the previous studies the scalar perturbations were considered in the hydrodynamical approach which works well for the early Universe.
It is obvious that the cosmological models should be consistent with the observations at all stages of the evolution of the Universe (both early and late).

Taking into account that the perturbations of the considered perfect fluids are purely adiabatic (i.e. dissipative processes are absent), we have shown that
such perfect fluids are compatible with the theory of scalar perturbations if they satisfy two conditions. First, these fluids must be clustered (i.e.
inhomogeneous). Second, the parameter of the equation of state $\omega$ should be $-1/3$. Therefore, this perfect fluid neither accelerates nor decelerates the
Universe. The frustrated network of the cosmic strings can be a candidate for this fluid. On the other hand, frustrated domain walls are ruled out because they
have\footnote{This result may change if we take into account the shear deformations of the perfect fluid. However, this problem is out of the scope of our
model and requires a separate investigation.} $\omega=-2/3$.

Therefore, in the case of negative constant $\omega$, only models with $\omega =-1$ (a pure cosmological constant) and $\omega =-1/3$ are compatible with the
mechanical approach, which is the most appropriate to describe the late Universe inside the cell of uniformity. Substituting $\omega =-1/3$ into the background
equation \rf{2.4a}, we can see that such perfect fluid behaves here as curvature. Hence, we can combine both terms to get a total "curvature" density parameter
$\Omega_{\mathcal{K},\mathrm{tot}}\equiv \Omega_{\mathcal{K}}+\Omega_{\mathrm{pf}}$. It is tempting to use the experimental restrictions on the curvature
density parameter (see, e.g., sections 4.3 and 6.2.3 in \cite{7WMAP} and \cite{Planck2013}, respectively) applying them for $\Omega_{\mathcal{K},\mathrm{tot}}$
and then to get limitations for $\lambda$ from Eq. \rf{3.3}. Exactly this parameter $\lambda$ determines the characteristic scales of the Yukawa-type screening
in formulae \rf{3.6} and \rf{3.12}.  However, we cannot do it because the experimental restrictions have topological origin (i.e. they are due to the different
form of the function $\Sigma$ in \rf{2.1a}) but not due to the fact that the curvature term in the Friedmann equations behaves as $1/a^2$. In other words, the
topological restrictions follow from the different definitions for the distances in the case of different topologies.

Then, we have obtained the equation for the nonrelativistic gravitational potential. We need to know the form of the gravitational potential to describe
dynamics of inhomogeneities. For example, all numerical simulations use the expression for the gravitational potentials of the inhomogeneities. Obviously,
dynamical behavior of these inhomogeneities is determined by two competing mechanisms. On the one hand, it is the gravitational interaction between the
inhomogeneities, and, on the other hand, the cosmological accelerated expansion. Therefore, one of the main tasks of the present paper was to study a
possibility to get a reasonable form of the gravitational potential in the considered model. We have shown that due to the perfect fluid with $\omega=-1/3$,
the physically reasonable solutions of the equation for the gravitational potential take place for flat, open and closed Universes. The presence of this
perfect fluid helps to resolve the Seeliger paradox \cite{Norton} for any sign of the spatial curvature parameter $\mathcal{K}$. If the perfect fluid is
absent, the hyperbolic space is preferred \cite{EZcosm1}. Hence, such perfect fluid can play an important role. This perfect fluid is concentrated around the
inhomogeneities and results in screening of the gravitational potential. It should be noted that the obtained gravitational potentials have an important
property: the total gravitational potentials averaged over the whole Universe are equal to zero $\overline\varphi=0$. Because the perfect fluid energy density
fluctuation is proportional to the total gravitational potential $\delta\varepsilon \sim \varphi$, then the averaged energy density fluctuation is also equal
to zero $\overline{\delta\varepsilon}=0$. Therefore, we arrive at the natural condition that the total perfect fluid energy density
$\varepsilon=\overline\varepsilon +\delta\varepsilon$ after the procedure of the averaging is equal to $\overline\varepsilon$.

It must be emphasized that the case of imperfect fluids with the varying parameter $\omega$ (e.g., scalar fields with arbitrary potentials) requires a separate
consideration which may lead to quite different conclusions. We intend to investigate this case in our forthcoming paper.

\section*{Acknowledgements}

The work of M. Eingorn was supported by NSF CREST award HRD-1345219 and NASA grant NNX09AV07A.


\end{document}